\documentclass[twocolumn,aps,prl,showpacs,superscriptaddress,unsortedaddress]{revtex4}
\usepackage{graphicx}
\usepackage{epsf}
\usepackage{color}
\newcommand{\etal}{{\it et al.}}

\begin{document}

\title{The coherent {\it d}-wave superconducting gap in underdoped
La$_{2-x}$Sr$_{x}$CuO$_4$ as studied by angle-resolved
photoemission}

\author{M. Shi}
\affiliation{Swiss Light Source, Paul Scherrer Institute, CH-5232 Villigen PSI, Switzerland}
\author{J. Chang}
\affiliation{Laboratory for Neutron Scattering, ETH Zurich and Paul
Scherrer Institute, CH-5232 Villigen PSI, Switzerland}
\author{S. Pailh\'es}
\affiliation{Laboratory for Neutron Scattering, ETH Zurich and Paul
Scherrer Institute, CH-5232 Villigen PSI, Switzerland}
\author{M. R. Norman}
\affiliation{Materials Science Division, Argonne National
Laboratory, Argonne, IL 60439 USA}
\author{J. C. Campuzano}
\affiliation{Department of Physics, University of Illinois at Chicago, Chicago, IL 60607 USA}
\affiliation{Materials Science Division, Argonne National Laboratory, Argonne, IL 60439 USA}
\author{M. M\aa{}nsson}
\affiliation{Materials Physics, Royal Institute of Technology KTH, S-164 40 Kista, Sweden}
\author{T. Claesson}
\affiliation{Materials Physics, Royal Institute of Technology KTH, S-164 40 Kista, Sweden}
\author{O. Tjernberg}
\affiliation{Materials Physics, Royal Institute of Technology KTH, S-164 40 Kista, Sweden}
\author{A. Bendounan}
\affiliation{Laboratory for Neutron Scattering, ETH Zurich and Paul
Scherrer Institute, CH-5232 Villigen PSI, Switzerland}
\author{L. Patthey}
\affiliation{Swiss Light Source, Paul Scherrer Institute, CH-5232 Villigen PSI, Switzerland}
\author{N. Momono}
\affiliation{Department of Physics, Hokkaido University Ð Sapporo 060-0810, Japan}
\author{M. Oda}
\affiliation{Department of Physics, Hokkaido University Ð Sapporo 060-0810, Japan}
\author{M. Ido}
\affiliation{Department of Physics, Hokkaido University Ð Sapporo 060-0810, Japan}
\author{C. Mudry}
\affiliation{Condensed Matter Theory Group, Paul Scherrer Institute, CH-5232 Villigen PSI, Switzerland}
\author{J. Mesot}
\affiliation{Laboratory for Neutron Scattering, ETH Zurich and Paul
Scherrer Institute, CH-5232 Villigen PSI, Switzerland}

\begin{abstract}
We present angle-resolved photoemission spectroscopy (ARPES) data on
moderately underdoped La$_{1.855}$Sr$_{0.145}$CuO$_4$ at
temperatures below and above the superconducting transition
temperature. Unlike previous studies of this material, we observe
sharp spectral peaks along the entire underlying Fermi surface in
the superconducting state. These peaks trace out an energy gap that
follows a simple {\it d}-wave form, with a maximum superconducting
gap of 14 meV. Our results are consistent with a single gap picture
for the cuprates. Furthermore our data on the even more underdoped
sample La$_{1.895}$Sr$_{0.105}$CuO$_4$ also show sharp spectral
peaks, even at the antinode, with a maximum superconducting gap of
26 meV.
\end{abstract}

\pacs{74.72.Dn, 74.25.Jb, 79.60.Bm}
\date{\today}
\maketitle

The energy gap is a fundamental property of
superconductors~\cite{BCS}.  The nature of its anisotropy has played
a key role in the testing and building of microscopic theories of
the superconductivity discovered in layered copper oxides~\cite{BM}.
This anisotropy can be measured by angle-resolved photoemission
spectroscopy (ARPES), which is a unique probe of electronic
excitations and their momentum dependence~\cite{JCRev,ZXRev}.
For underdoped samples, an energy gap persists above {\it
T}$_c$~\cite{Ding,Loeser}. This pseudogap is most prominent
in the $(\pi,0)$ region of the Brillouin zone, giving rise to a
gapless arc of states centered about the zone diagonal known as a
Fermi arc~\cite{Nat98}. How the gap evolves from the superconducting
state to the pseudogap phase remains one of the most important
questions being debated in the cuprates~\cite{NPK}. It has been
suggested that the superconducting gap only exists on the Fermi
arcs, and thus is distinct from the pseudogap~\cite{Hashimoto}. This
``two gap'' scenario has been supported by recent ARPES studies on
optimally doped La$_{2-x}$Sr$_x$CuO$_4$ (LSCO)~\cite{Terashima} and
Bi$_2$Sr$_2$CuO$_6$ (Bi2201)~\cite{Kondo}, as well as on heavily
underdoped Bi2212~\cite{Tanaka}. In these studies,
sharp spectral peaks are observed  below {\it T}$_c$ only along the arc, whereas the
states in the  region around the $(\pi,0)-(\pi,\pi)$, Fermi crossing
(the antinode) remain incoherent, as they were above {\it T}$_c$.
Moreover, the size of the energy gap in the antinodal region is
significantly larger than that expected from a simple extrapolation
of the gap from the arc.

Our ARPES results on moderately underdoped LSCO ($x = 0.145$) reveal
a very different picture. Similar to previous studies of underdoped
Bi2212~\cite{Amit}, below {\it T}$_c$ we observe sharp spectral
peaks along the entire underlying Fermi surface (FS), which trace out a
simple {\it d}-wave gap with a maximum amplitude of 14~meV. For more
underdoped LSCO ($x = 0.105$) the spectra are still characterized by
coherent peaks in the $(\pi,0)$ region with a maximal gap amplitude
of 26~meV. We see no evidence for a much larger gap in the antinodal
region as reported in other studies~\cite{Tanaka,Kondo,Terashima}.
More significantly, we find that the superconducting and pseudogaps
have similar maximum amplitudes.

\begin{figure*}
\includegraphics
[width=6.0in]
 {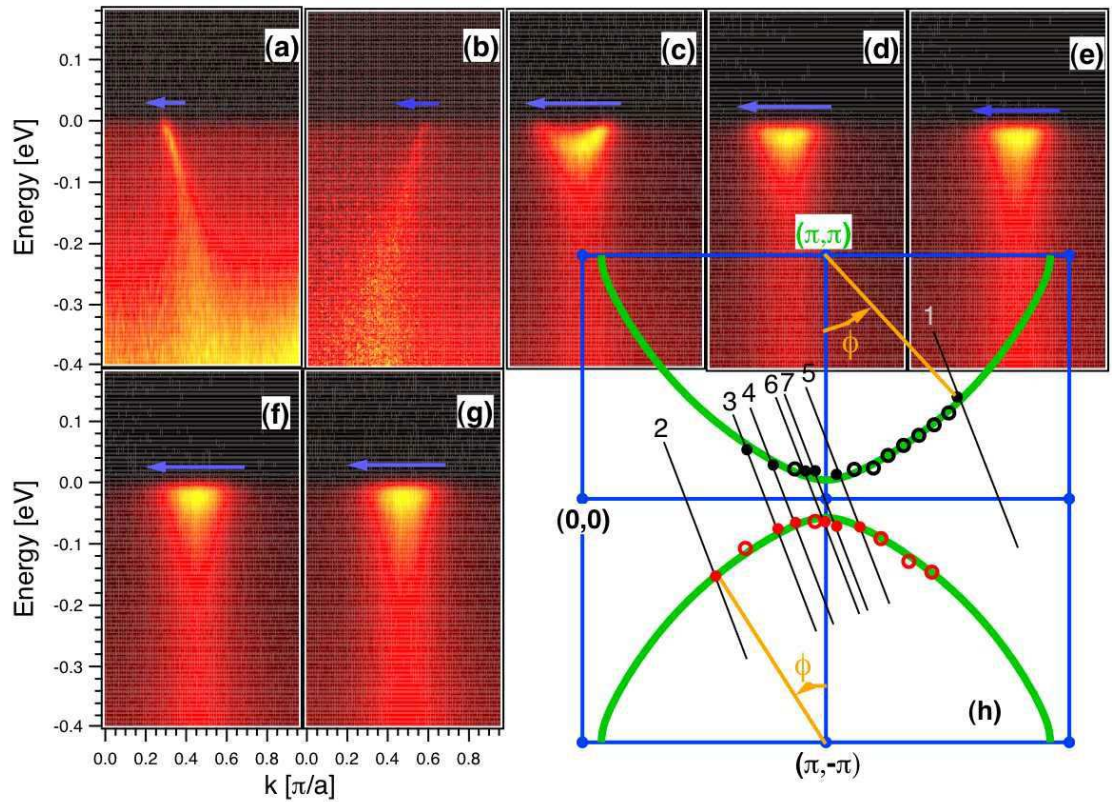}
\caption{(Color online) ARPES intensities along momentum cuts that cross the
underlying Fermi surface (FS) (a) at the node, (b) to (f) between the node and the
antinode, and (g) at the antinode. The corresponding cuts 1 -- 7 are
indicated in (h). The circles in (h) are the underlying FS
determined from our measurements, while the solid line is the FS
from a tight-binding fit to our data. The measurements were
performed on LSCO ($T_c=36$ K, $x = 0.145$) at 12~K with {\it h}$\nu$ = 55 eV.}
\label{fig1}
\end{figure*}

ARPES results were obtained on very high quality single crystals of
LSCO, grown using the travelling solvent floating zone method
\cite{Nakano}, with a transition temperature $T_c$ = 36~K and 30~K
for $x = 0.145$ and $x = 0.105$, respectively. The transition width
for both dopings are $\Delta T_c\approx$ 1.5~K. ARPES experiments
were carried out at the Surface and Interface Spectroscopy beamline
at the Swiss Light Source. During the measurements, the base
pressure always remained less than 5 $\times$ 10$^{-11}$ mbar. The
ARPES spectra were recorded with a Scienta SES2002 electron analyzer
with an angular resolution of better than 0.15$^\circ$. Circularly
polarized light with {\it h}$\nu$ = 55 eV and linearly polarized
light with {\it h}$\nu$ = 25 eV were used. The energy resolutions
were 17 meV and 12 meV for {\it h}$\nu$ = 55 eV and {\it h}$\nu$ =
25 eV, respectively. The Fermi level was determined by recording the
photoemission spectra from polycrystalline copper on the sample
holder. The samples were cleaved {\it in situ} by using a specially
designed cleaving tool~\cite{Mansson}. Clear (1$\times$1) low-energy
electron-diffraction patterns obtained after the ARPES measurements
indicate that the cleaved surfaces had good quality.

Figure~1 shows typical ARPES intensity maps of LSCO ($x = 0.145$) in
the superconducting phase as a function of wavevector and energy in
the nodal region [Fig.~1(a)], in the antinodal region [Figs.~1(f)
and 1(g)], and between these two regions [Figs. 1(b) to 1(e)]. While
the spectral peaks close to the nodal direction exhibit a sizable
dispersion, the dispersion is much weaker in the antinodal region,
which is the ubiquitous behavior among hole-doped cuprates in the
superconducting state. We used two different methods to determine
the underlying normal state FS. In the first case, it was
obtained from the peak positions of the momentum distribution curves
at zero binding energy along momentum cuts parallel to those shown
in Fig.~1(h), and in the second method, it was identified by
searching for the minimum gap location of the energy distribution
curves (EDCs) along each cut. Both methods gave the same result.
From a tight-binding fit to the underlying FS, we obtain a hole
volume of 57.2\%, corresponding to 1.144 holes per Cu atom, in
agreement with the Sr doping level of $x=0.145$.

It is important to mention that we have observed sharp spectral
peaks along the {\it entire} underlying FS in the superconducting
state. To visualize this, in Fig.~2  we plot EDCs along the momentum
cuts 1, 3 and 7 in Fig.~1(h) in reduced momentum ranges. The
corresponding momentum windows are marked in Figs.~1(a), 1(c), and
1(g) with arrows. A comparison of the spectrum at the node and the
antinode is given in Fig.~2(d). One can see both spectra are
characterized by a sharp peak with a similar width. The true widths
of these peaks are even smaller, given the energy resolution (17
meV) used in these measurements. To demonstrate this, based on the
observation that the spectral peaks disperse only weakly in the
$(\pi,0)$ region, we collected an angle-integrated EDC in the
momentum window marked in Fig.~1(g) by the arrow with a higher
energy resolution of 12 meV. Indeed, as expected, the peak narrows
in energy as can be seen in Fig.~2(e), where we compare this
angle-integrated EDC with the EDC at the antinodal {\it k}$_F$
acquired with a resolution of 17 meV. The sharp spectral peaks we
observe near $(\pi,0)$ are very different from the broad EDCs
reported in most ARPES studies of LSCO.

\begin{figure}
\includegraphics
[width=\columnwidth]
 {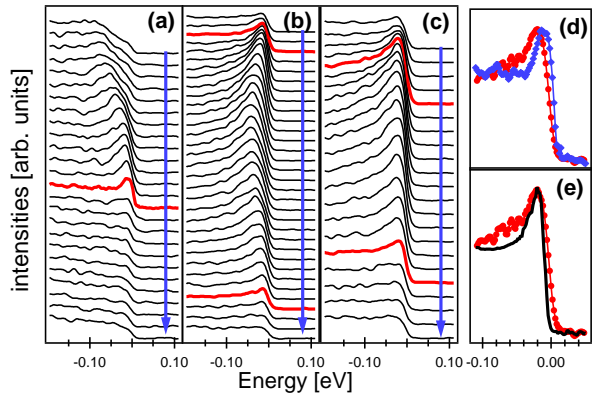}
\caption{(Color online) (a) -- (c) EDCs along the momentum cuts 1, 3
and 7 in Fig.~1(h). The corresponding momentum ranges are indicated
in Figs.~1(a), 1(c), and 1(g) with arrows. Thick curves are EDCs at {\it
k}$_F$. (d) Comparison of the EDCs at the antinode (left curve) and
at the node (right curve). (e) Comparison of the EDC at the antinode
with {\it h}$\nu$ = 55 eV and an energy resolution of 17 meV (solid
symbols), and the angle-integrated EDC over the momentum range
marked by the arrow in Fig.~1(g) acquired with {\it h}$\nu$ = 25 eV
an energy resolution of 12 meV (solid line).} \label{fig2}
\end{figure}

To determine the superconducting gap, we ``symmetrize'' the EDCs at
{\it k}$_F$ to effectively eliminate the Fermi function from the
measured ARPES spectra~\cite{Nat98}. At the antinode, the
symmetrized EDC shows two sharp peaks with a clearly defined gap,
EDC 7 in Fig.~3(a). Moving from the antinode towards the node along
the underlying FS, the separation between the two peaks
becomes smaller, and the spectral weight at zero energy fills in,
but a two-peak structure is always evident, even for EDC 3 where the
gap and spectral width are comparable. Closer to the node, the
spectral gap continues to decrease, and finally, when the node is
reached, a single peak in the symmetrized EDC is seen (EDC 1).

\begin{figure*}
\includegraphics
[width=6.0in]
 {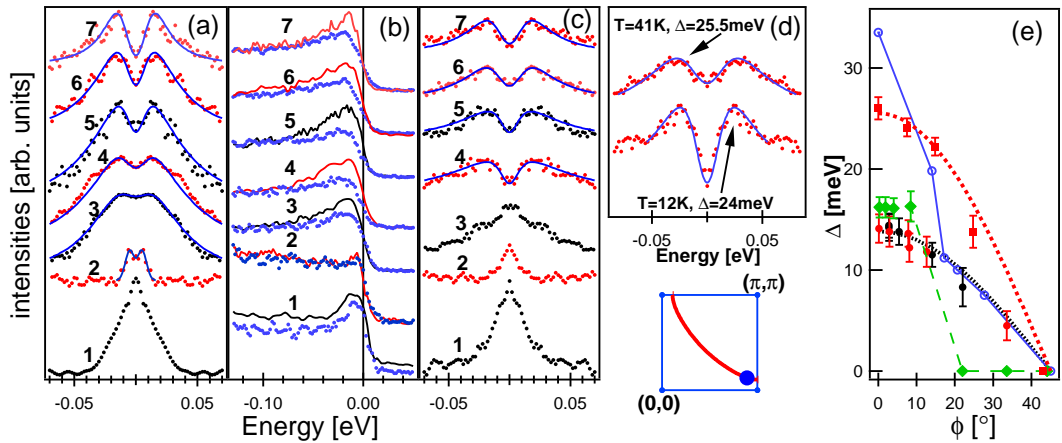}
\caption{(Color online) (a) and (c) Symmetrized EDCs for $x = 0.145$
at various {\it k}$_F$ from the antinode (top) to the node (bottom)
(each curve is offset for clarity) at 12 K and 41 K, respectively.
The corresponding {\it k}$_F$ from cuts 1 -- 7 are indicated in
Fig.~1(h) by filled circles. The {\it k}$_F$ for cuts 1, 3, and 5
are on the upper underlying FS curve, and the rest are on the lower
underlying FS curve.
The solid lines are fits using Eqs.~(1) and (2). (b) The EDCs at 12
K (solid lines) and at 41 K (dotted lines) from which the
symmetrized EDCs in (a) and (c) are obtained. (d) Symmetrized EDCs
for $x = 0.105$ at the {\it k}$_F$ shown below the panel. (e) The
superconducting gaps for $x = 0.105$ (filled squares), the
superconducting gaps (filled circles) and the pseudogaps for $x =
0.145$ (diamonds connected with a dashed line) from the fit as a
function of the underlying FS angle $\phi$ indicated in Fig.~1(h),
with the dotted lines the simple {\it d}-wave gap
$\Delta_{max}|\cos(2\phi)|$ with $\Delta_{max}$ = 13.8 meV and 25.5
meV, respectively. The open circles are the gaps determined by
Terashima \etal~for a LSCO sample with $x = 0.15$~\cite{Terashima}.
The data were acquired with {\it h}$\nu$ = 55 eV.} \label{fig3}
\end{figure*}

To perform a more quantitative analysis of the superconducting gap,
we use a simple form for the self-energy that has proven successful
in modeling ARPES data~\cite{PRB98}
\begin{equation}
\Sigma(k_F,\omega) = -i\Gamma_1 + \frac{\Delta^2}{\omega+i\Gamma_0}
\end{equation}
which is then used to calculate the spectral function
\begin{equation}
A(k_F,\omega) =
\frac{1}{\pi}
\frac{\mathrm{Im}\,\Sigma}{(\omega-\mathrm{Re}\,\Sigma)^2+(\mathrm{Im}\,\Sigma)^2}
\end{equation}
This is then convolved with the experimental resolution and fitted
to the symmetrized EDC. Here $\Gamma_1$ is the single-particle
scattering rate, and $\Gamma_0$ is related to the inverse pair
lifetime. The fitted curves are plotted as solid lines in Fig.~3(a).
The best fit was achieved at the antinode where the separation of
the two peaks is the largest, with $\Delta$ = 13.8 $\pm$ 1.5 meV,
$\Gamma_1$ = 38 $\pm$ 3 meV, and $\Gamma_0$ is negligibly small
(less than 0.1 meV). Near the node, we find $\Gamma_1\sim 9$~meV, a
scattering rate that compares well with the one estimated from the
recent infrared conductivity measurements on
La$_{1.83}$Sr$_{0.17}$CuO$_4$ in Ref.~\cite{optics}. The negligibly
small value of $\Gamma_0$ indicates that the gap at the antinode is
associated with pairs which have infinite lifetime (long range
order) in the superconducting state. Moreover, the variation of the
amplitude of the gap follows a simple {\it d}-wave form over the
entire FS, Fig.~3(e).

We now address the issue about the existence of the pseudogap above
{\it T}$_c$ in LSCO ($x = 0.145$) samples. In Fig.~3(b) we plot the
EDCs taken at 41~K along the underlying FS and compare them with
those taken at 12~K. While at the node the EDC at 41 K is still
characterized by a sharp spectral peak, the EDCs for the antinodal
region are characterized by broad peaks with a suppression of
spectral weight relative to those below {\it T}$_c$. The symmetrized
EDCs (Fig.~3c) clearly show that above {\it T}$_c$ there is a gap in
the electronic excitation spectra in the antinodal region and there
is a Fermi arc along which the excitation spectra are gapless. The
maximal size of the pseudogap as extracted from the fits by using
Eqs (1) and (2) has a similar amplitude to that of the
superconducting gap [Fig.~3(e)]. It should be noted that the
determined pseudogap is much smaller than that reported in
Refs~\cite{Yoshida,Terashima}.

Our ARPES results on LSCO  for $x = 0.145$ are consistent with one
gap in the superconducting state, but not with a gap having two
components, namely, a superconducting gap along the Fermi arc and a
pseudogap in the antinodal region with a higher energy scale as seen
in previous photoemission studies of LSCO~\cite{Yoshida,Terashima},
as well as underdoped Bi2212~\cite{Tanaka} and optimally doped
Bi2201~\cite{Kondo}. The first difference is that we observe sharp
spectral peaks along the entire underlying FS, while
spectra in these other studies are broad in the antinodal region.
The second difference concerns the momentum dependence of the energy
gap. In the arc region, the gaps from all of these studies follow a
simple {\it d}-wave form. But in the antinodal region, the gaps in
the other ARPES studies cross over to a abnormally large pseudogap
at the same location where the sharp spectral peak disappears.  But
in our case, the sharp spectral peak remains and the gap continues
to follow the simple {\it d}-wave form, with a maximum value of 14
meV at the antinode.

The one gap nature of underdoped cuprates is further supported by
our ARPES results on more underdoped LSCO ($x = 0.105$) which, as
for the $x = 0.145$ samples, reveal: (1) the symmetrized EDCs near
the antinode below (above) {\it T}$_c$ are characterized by sharp
(broad) peaks [Fig.~3(d)], (2) the maximal gap in the electronic
excitation spectra below and above {\it T}$_c$ has a similar
amplitude, and (3) the anisotropy of the energy gap in the
superconducting state is consistent with a {\it d}-wave form
[Fig.~3(e)], though more data will be needed near the node in order
to determine whether any gap flattening is present as observed in
underdoped Bi2212~\cite{MesotPRL99}. In addition, we find that the
maximal gap for $x = 0.105$ is almost twice that of $x = 0.145$.
This increase in the energy gap with underdoping follows the same
trend as previously observed in Bi2212~\cite{JCRev, ZXRev,
MesotPRL99}.

From the momentum dependence of the gap and the observation that
this gap is always associated with a sharp spectral peak, we
conclude that the gap we measured below {\it T}$_c$ is associated
with the {\it d}-wave superconducting order parameter. Several
scenarios can be put forward to explain the difference between our
data and those reported earlier. One possibility may be due to
increased scattering in the antinodal region resulting from disorder
or imperfection that, depending on the sample quality, varies in
magnitude. The fact that the samples used for our study appear to be
of superior quality is confirmed by two neutron scattering
experiments performed on the same sample~\cite{Chang}, where (1) a
clear vortex lattice could be measured, and (2) a clean spin gap
could be observed below {\it T}$_c$, which, in LSCO, is known to be
visible only in high-quality samples. Several factors can affect the
quality of the samples and, consequently of the ARPES spectra. Of
crucial importance is the sample preparation, but other factors such
as how the samples are cleaved can play an important role. We have
noticed that the specially designed cleaver mentioned
above~\cite{Mansson} enables us to obtain data of higher quality
than that obtained using the very popular ``post and glue" cleaving
method. Finally, it is interesting to notice that the weak (strong)
sensitivity of the nodal (antinodal) states to disorder is
consistent with scanning tunneling spectroscopy (STM) studies of
Bi2212~\cite{McElroy}. In these STM studies, spectra for biases
comparable to that of the antinodal states showed strong sensitivity
to spatial location, accompanied by a large distribution of gap
sizes.  For low biases, though, with energies comparable to states
near the node, the spectra are homogenous.

In summary, we have measured the superconducting gap and its
momentum dependence by ARPES in moderately underdoped LSCO at
temperatures well below {\it T}$_c$.  For $x = 0.145$ the anisotropy
of the gap follows a simple {\it d}-wave form, similar to that
observed in optimally and overdoped Bi2212. We find no evidence for
the coexistence of a superconducting gap and a pseudogap, and relate
this finding to the presence of sharp spectral peaks along the
entire underlying FS. Such sharp features at the nodal
and antinodal points remain even at the lower doping level of $x =
0.105$, although more data closer to the nodal point will be needed
in order to determine the precise momentum dependence of the
$d$-wave gap in this case. Whether our one-gap scenario continues to
hold for more underdoped LSCO samples will be an important matter
for future studies.

This work was supported by the Swiss National Science Foundation
(through NCCR-MaNEP, and grant No. 200020-105151), the Ministry of
Education and Science of Japan, the Swedish Research Council, and
the U.S. DOE, Office of Science, under Contract
No.~DE-AC02-06CH11357 and by NSF DMR-0606255. This work was
performed at SLS of the Paul Scherrer Insitut, Villigen PSI,
Switzerland. We thank the beamline staff of X09LA for their
excellent support.

\end{document}